\documentclass[conference]{IEEEtran}
\IEEEoverridecommandlockouts
\usepackage[utf8]{inputenc}
\usepackage{graphicx}
\usepackage{mathrsfs}
\usepackage{mathtools}  
\usepackage{amsfonts}  
\usepackage{amsmath}
\usepackage{algorithm}
\usepackage{multicol}
\usepackage{textcomp}
\usepackage{bbm}
\usepackage{comment}
\usepackage{tikz}
\usepackage{xcolor}
\usetikzlibrary{positioning, arrows, shapes, calc}
\usepackage{subcaption} 
\usepackage{balance}

\usepackage{svg}

\usepackage{algpseudocode}
\usepackage[colorinlistoftodos]{todonotes}

\newcommand{\thetaa}{\theta_{\text{actor}}}
\newcommand{\thetac}{\theta_{\text{critic}}}
\newcommand{\cexpert}{{\mathcal{C}}_{\text{expert}}}

\title{Adaptive Design of mmWave Initial Access Codebooks using Reinforcement Learning}

\author{\normalsize
Sabrine Aroua$^*$, Christos A. Bovolis$^*$, Bo Göransson$^\dag$, Anastasios Giovanidis$^*$, Mathieu Leconte$^*$, Apostolos Destounis$^*$ \\$^*$Ericsson Research, Massy, France\\
$^\dag$Ericsson, Stockholm, Sweden}

\begin{document}

\newcommand{\hif}[1]{{\color{red}~{\bf [Hassan:} #1]~}}

\maketitle

\begin{abstract}
Initial access (IA) is the process by which user equipment (UE) establishes its first connection with a base station. In 5G systems, particularly at millimeter-wave frequencies, IA integrates beam management to support highly directional transmissions. The base station employs a codebook of beams for the transmission of Synchronization Signal Blocks (SSBs), which are periodically swept to detect and connect users. The design of this SSB codebook is critical for ensuring reliable, wide-area coverage.
In current networks, SSB codebooks are meticulously engineered by domain experts. While these expert-defined codebooks provide a robust baseline, they lack flexibility in dynamic or heterogeneous environments where user distributions vary, limiting their overall effectiveness.
This paper proposes a hybrid Reinforcement Learning (RL) framework for adaptive SSB codebook design. Building on top of expert knowledge, the RL agent leverages a pool of expert-designed SSB beams and learns to adaptively select or combine them based on real-time feedback. This enables the agent to dynamically tailor codebooks to the actual environment, without requiring explicit user location information, while always respecting practical beam constraints.
Simulation results demonstrate that, on average, the proposed approach improves user connectivity by 10.8$\%$ compared to static expert configurations.
These findings highlight the potential of combining expert knowledge with data-driven optimization to achieve more intelligent, flexible, and resilient beam management in next-generation wireless networks.

\end{abstract}

\begin{IEEEkeywords}
Initial access, SSB codebook design, Expert-designed
beams, Reinforcement Learning (RL).
\end{IEEEkeywords}

\section{Introduction}
\label{sec:introduction}
In 5G, millimeter wave (mmWave), also referred to as FR2, covers frequencies starting from around 24 GHz up to 40 GHz in current commercial deployments, with the full FR2 range extending up to 71 GHz. These bands provide access to very large amounts of spectrum, with allocations often reaching 800 MHz or more per operator per band, enabling extremely high data rates and capacity. Owing to these characteristics, mmWave/FR2 is considered a cornerstone for dense urban areas, stadiums, and hotspot scenarios, where traffic demand exceeds the capacity of sub-6 GHz (FR1) networks \cite{ER_mmWave}.

However, the use of FR2 also introduces new challenges in coverage and in maintaining reliable communication with User Equipments (UEs). At such high frequencies, the shorter wavelength enables the deployment of large antenna arrays that generate narrow beams, providing the array gain necessary to overcome the severe path loss \cite{raghavan2016beamforming}. This makes beam management, the process of establishing, maintaining and refining directional links, a fundamental aspect of 5G operation at mmWave. The first stage of beam management is the Initial Access (IA) procedure, by which a UE establishes a connection to a suitable cell. IA relies on Synchronization Signal Blocks (SSBs). While in FR1 a single wide beam is often sufficient to broadcast an SSB, in FR2 the narrow beams cannot cover an entire cell sector. Consequently, multiple SSBs are transmitted in different beam directions. The 3GPP standard allows up to $64$ SSBs per cell, ensuring full coverage so that nearly all UEs can detect and synchronize with the network \cite{3gpp_tr_21_917}.

\begin{figure}[h]
\begin{center}
    \includegraphics
    {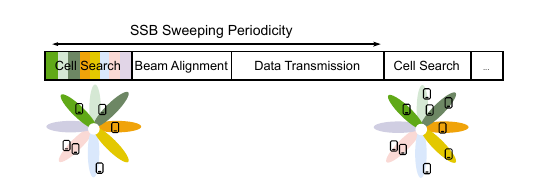}   
    \caption{Cell Search (CS) procedure.}
    \label{fig:perf_fig1}
 \end{center}  
\end{figure}

In 5G, the Initial Access (IA) procedure typically consists of two main stages: Cell Search (CS) and Beam Alignment (BA) \cite{BF2018tutorial}. During the CS phase, the cell sweeps the surrounding environment using a predefined codebook of SSB beams \cite{barati2016initial}. The UE listens to these beams and selects the cell whose beam provides the highest received power, as long as it exceeds the minimum threshold required for association. This initial connection triggers the BA phase, during which both the cell and the UE refine their transmit–receive beam pair to establish a reliable directional link \cite{tandler2023deep}. Afterward, as illustrated in Figure \ref{fig:perf_fig1}, data transmission begins. The SSB beam sweeping periodicity typically ranges from $5$ to $120$ ms, depending on the configuration and beamforming strategy. \cite{3gpp_ts_19}.

The design of the SSB codebook is crucial to enabling reliable communication, as it directly impacts cell coverage, detection accuracy, and initial access performance. Traditionally, these codebooks are designed manually by radio domain experts, who define the spatial coverage strategy during the CS phase.
Despite their meticulous design, expert-crafted SSB codebooks face inherent limitations due to their reliance on fixed heuristics and static propagation assumptions. Rigid beam patterns cannot adapt to dynamic environments, resulting in suboptimal coverage when faced with mobility, blockages, or irregular user distributions. Moreover, while manual designs account for known deployment scenarios, experts may overlook emerging or unconventional use cases that deviate from the initial assumptions. 
These limitations highlight the need for more adaptive and data-driven approaches. Artificial Intelligence (AI), and in particular Reinforcement Learning (RL), offers a promising alternative by enabling the dynamic optimization of codebooks based on real-time network conditions. Unlike expert-designed solutions, RL can learn from the environment, adjust to changing UE distributions and propagation characteristics, and continuously improve beam selection strategies to enhance IA performance.

\subsection{Prior Work} \label{sec:RW}

Recent advances have leveraged data-driven techniques to improve SSB codebook design, moving beyond traditional grid-of-beams and statistical approaches \cite{li2017design, giordani2016initial}.

In \cite{dreifuerst2023ml, dreifuerst2024hierarchical}, supervised learning is used to jointly design SSB and CSI-RS codebooks for sub-6 GHz 5G NR. Neural networks are trained offline on channel datasets to generate site-specific codebooks, which replace expert designs until retraining. Their performance is benchmarked against Discrete Fourier Transform (DFT) beamformers that provide uniform coverage with narrow or wide beams. In contrast, \cite{zhang2021reinforcement} proposes an RL approach for mmWave and Terahertz networks, using only received power measurements from the UE side for each SSB beam swept. Users are clustered based on the SSB feedback (hereafter referred to as the sensing-beam feedback), and each cluster is managed by a dedicated deep RL agent.
Then, using a Wolpertinger-based \cite{wolpertingerArchitecture} latent action space, agents learn beam patterns that maximize the average SNR, collectively forming the overall codebook. These approaches operate in two phases: Learning and Deployment. During the Learning phase, the codebook is trained with minimal impact on system performance. Once learned, it replaces the static codebook for deployment. However, like expert-designed codebooks, the learned codebook cannot adapt to unexpected traffic or environmental changes, and any update or replacement requires a new training phase.

The work in \cite{che2025efficient} addresses this limitation by using RL to dynamically select subsets of SSBs during each IA period. The objective is to reduce beam-sweeping time while maintaining cell discovery performance. In this approach, the RL policy selects a group of SSB beams from a reduced codebook to associate users, potentially updating the group every $20$ ms, which corresponds to the default SSB periodicity \cite{3gpp_ts_19}.

While this method introduces adaptability by enabling different SSB groups to be used over time, the temporal stability of the SSB sweep pattern remains crucial. Because SSBs are tied to beam management, mobility, and control signaling, frequent reconfiguration (e.g., every tens of milliseconds) could disrupt synchronization, handover, and beam tracking procedures expected by UEs.

Therefore, the flexibility lies not in changing the SSB set every IA burst, but rather in adapting the SSB configuration over longer time scales (e.g., seconds or minutes) or across different spatial sectors based on traffic dynamics or interference conditions. This preserves the periodicity and structure required for reliable operation while allowing slow, context-aware adaptation of the SSB sweeping strategy.

\subsection{Our Contribution} \label{sec:RW}

For the rest of this paper, the terms "SSB," "beam," and "precoding vector" are used interchangeably.

In this paper, we propose an RL-based solution to learn the SSB codebook for IA using precoding vectors already designed by domain experts. Typically, experts design multiple SSB groups, each tailored to specific traffic distributions, and deploy them using static heuristics.
For example, one codebook (or group of SSBs) may be optimal for an open square, whereas another may better suit a train station or university campus. Such static selection, however, cannot adapt to dynamic or unforeseen changes in user distribution or environmental conditions.

To overcome this limitation, the RL agent accesses a dataset of expert-designed SSBs and learns to adaptively select the most suitable beam group at runtime. Importantly, it is not restricted to predefined groups but can form new combinations from the expert set, expanding the design space beyond static heuristics. By leveraging feedback on user discovery performance, the agent dynamically chooses the set of SSBs that maximizes the number of discovered users, replacing static rules with a data-driven, adaptive policy. This approach bridges expert knowledge and real-time adaptability while ensuring safe exploration, as all beams originate from the expert-designed set. The learned codebook is kept fixed for a given period, for example, several minutes, to maintain the predictability and stability of SSB sweeping, but can be updated periodically to respond to changing network conditions. The RL agent guarantees at least the performance of expert-designed codebooks and may surpass it by exploring novel beam groupings. To our knowledge, this is the first work that benchmarks AI-learned SSB codebooks against expert codebooks.

The remainder of the paper is organized as follows. Section \ref{sec:model}, introduces the system model and defines the adaptive SSB codebook design problem. Section \ref{sec:RL} presents the RL-based solution and beam design methodology. Section \ref{sec:perf} details the simulation setup and reports numerical results comparing learned and expert codebooks. Section \ref{sec:conc} concludes the paper and outlines potential future research directions.


\section{Network Model and Problem Statement}
 \label{sec:model}
 This section presents the network and beamforming model for mmWave IA.
It also defines the SSB codebook design problem as selecting SSBs to maximize user coverage and traffic offloading.

\subsection{System Model}

We consider a system model in which a mmWave massive MIMO Base Station (BS) serves multiple single-antenna users. The BS is divided into multiple sectors $\mathcal S$ (typically three), each equipped with a planar dual-polarized antenna array of $e_1 \times e_2$ elements along the elevation and azimuth dimensions. The array supports variable amplitude and phase shifters, enabling adaptable beamwidths in both domains. Each sector is equipped with a set of expert-designed SSB codebooks $\cexpert$, where each codebook $c \in \cexpert$ is a fixed subset containing exactly $n$ SSB beams:  

\[
c = \{w_1^c, w_2^c, \ldots, w_n^c\}, \quad \forall c \in \cexpert.
\]  
Here, $w_b^c$ is the precoding beamforming vector for the $b$-th SSB in codebook $c$. It consists of $2 \times e_1 \times e_2$ complex weights, where the factor 2 accounts for dual polarization.  

Traditionally, during the CS phase, each sector $s\in\mathcal S$ sequentially broadcasts the SSB signal at the $n$ beams that comprise one of its predefined SSB codebooks $c_s \in \cexpert$. This codebook is selected by an expert and remains fixed for a given period.

Each user $u$ measures the signal power received from each beam transmitted during the CS phase, as in \cite{che2023efficient}. Assuming successful decoding, the power $P_u^{(s,w)}$ received by user $u$ from beam with precoding vector $w$ of sector $s$ is given by:
\begin{equation}
P_u^{(s,w)} = \|\mathbf{h}_{s,u}^T w~x\|,
\end{equation}
$\mathbf{h}_{s,u}$ denotes the downlink channel vector between sector $s$ and user $u$. $x$ represents the broadcast reference signals.

Based on these measurements, the user identifies the sector-beam pair that provides the maximum received power:  

\begin{equation}
(s^*_u, w^*_u) = \arg\max_{s \in \mathcal S,\; w \in c_s} P_{u}^{(s,w)}.
\end{equation}
The user is associated with sector $s^*_u$ and beam $w^*_u$ only if the received power exceeds the detection threshold $\tau$:

\begin{equation}
P_{u}^{(s^*_u, w^*_u)} \geq \tau.
\end{equation}  
Otherwise, the user remains unassociated, ensuring reliable communication only under adequate signal quality.

For each traffic distribution, the design of SSB precoding vectors is critical, as it determines the received power and, consequently, the probability of successful detection and association. Suboptimal designs can lead to weak signals or high interference, reducing coverage and impairing IA performance.

\subsection{Problem Definition}
In line with the CS procedure described above, we define $\mathcal B\subseteq\mathbb C^{2\times e_1\times e_2}$, with $|\mathcal B|=m$, as the full pool of expert-designed SSBs shared across all sectors $\mathcal S$. The predefined codebooks form only a subset of this pool, i.e., $w\in\mathcal B$ for all $w\in c\in\cexpert$. Consequently, $\mathcal B$ also contains beams not currently used in any common codebook. Such beams may be designed for different environments, deployment areas, or traffic distributions, or may come from inactive codebooks.

Building on this, we propose a solution in which each sector $s\in\mathcal S$ dynamically designs a new SSB codebook tailored to the prevailing traffic distribution. The admissible codebooks are the subsets of $\mathcal B$ of size $n$ and we let $\mathcal C=\left\{c \subseteq \mathcal B: |c|=n\right\}\supseteq\cexpert$. This adaptive codebook will be selected to maximize the number of users, i.e., wireless devices, that can successfully associate during CS. 
By leveraging the full beam pool $\mathcal B$, including those not used in the common codebooks, each sector gains access to a richer set of design choices. This enables the formation of adaptive codebooks that remain effective under dynamic or unforeseen conditions and improve user association performance.

A user is associated with sector $s\in\mathcal S$ if the received power from the strongest beam in $c_s$ exceeds threshold $\tau$. The user is considered associated if this holds for any sector. The objective is to maximize the expected number of associated users over all user distributions and channel realizations. Formally, the beam selection problem is:

\begin{eqnarray}
\max_{\forall s\in\mathcal S, c_s\in\mathcal C} \quad
\mathbb{E}\left[ \sum_{u} \mathbbm{1}\!\left\{ \max_{s\in\mathcal S, w\in c_s} P_u^{(s,w)} \geq \tau \right\} \right]\label{eqn: joint objective}\\
= \max_{\forall s\in\mathcal S, c_s\in\mathcal C} \quad \sum_{s\in\mathcal S}\mathbb E\left[\sum_{u: s^*_u=s} \mathbbm{1}\!\left\{ P_u^{(s^*_u,w^*_u)} \geq \tau\right\}\right]\label{eqn: separate objective}
\label{eq:beam_selection}
\end{eqnarray}

The indicator function $\mathbbm{1}\{\cdot\}$ equals $1$ if the condition inside the braces is satisfied (e.g., if user $u$ receives power above $\tau$ from any SSB) and $0$ otherwise. Thus, the summation in Equation~\eqref{eqn: joint objective} counts users for whom at least one SSB provides sufficient received power.
To limit complexity, each sector selects its codebook independently, though a user can be associated with only one sector. Hence, we omit the sector index when it is clear from context.

While alternative objectives, such as maximizing traffic offloading or serving users with strict QoS requirements, could be considered, in this work we focus on coverage, measured as the number of wireless devices successfully connected. This objective aligns with the key purpose of FR2 deployment, where ensuring reliable IA is a prerequisite for any traffic delivery or QoS guarantees. Importantly, the objective function is submodular, meaning that the incremental gain from adding a new beam diminishes as more beams are already selected. This property makes the problem equivalent to the Maximum Coverage Problem (MCP) and enables greedy algorithms to provide a $(1 - 1/e)$-approximation guarantee \cite{ageev1999approximation}.

Greedy methods rely on prior knowledge of user distributions and channel statistics, which are uncertain and time-varying. Reinforcement Learning (RL) offers a data-driven alternative, enabling the BS to learn adaptive beam-selection policies that respond to real traffic and channel variations.

\section{Adaptive Codebook Design with RL}\label{sec:RL}
Building on the problem formulation introduced in Section~\ref{sec:model}, we now present our solution framework. The proposed method leverages RL to address the combinatorial and information-constrained nature of codebook selection. The overall procedure consists of three main stages:

\begin{enumerate}
\item \textbf{SSB scanning with expert codebooks:} An initial sweeping is performed using the expert codebooks $\cexpert$ to collect sufficient observations of the network state. This sweeping can be extended over multiple cycles to ensure reliable measurements for all codebooks.
\item \textbf{SSB selection:} The collected measurements are provided as input to a neural policy network, which outputs the $n$ SSBs forming the selected codebook.
\item \textbf{Codebook deployment:} The selected codebook is deployed and used during subsequent CS periods. 
\end{enumerate}

When a new codebook is deployed, measurement collection continues, thus triggering a return to Steps 1--3
for a new codebook design. These measurements serve two purposes: (i) they allow the RL agent to update its policy parameters, enabling continual adaptation to network dynamics. and (ii) they enable the operator or expert system to determine when a new reconfiguration is needed.

To formalize this approach, we cast codebook selection as a Partially Observable Markov Decision Process (POMDP) \cite{egorov2015deep}. In this formulation, the true network state, which consists of the user locations and channel conditions as well as actions and measurements from the other sectors, is hidden from the sector optimizing its codebook, and decisions must be made based solely on partial observations.

\subsection{POMDP Formulation}

For a POMDP, it is necessary to define the action space, the observation space, and the reward function, which together specify the interaction between the agent and the environment:

\begin{itemize}
\item \textbf{Actions:} Selection of $n$ SSBs from a pool $\mathcal B$ of candidate expert-designed SSBs to deploy in the next CS.
\item \textbf{Observations:} The observation, obtained from Step 1, is a vector containing the number of UEs associated with each SSB across all beams and all codebooks in the expert set $\cexpert$, capturing the spatial distribution of UEs. For a sector $s\in\mathcal S$, codebook $c_s \in \cexpert$, and beam $w \in c_s$, the per-beam user count is
\begin{equation}
    o_{s,w} = \sum_{u: (s^*_u,w^*_u)=(s,w)} \mathbbm{1}\!\left\{ P_u^{(s^*_u,w^*_u)} \geq \tau\right\}.
\end{equation}
The full observation, $o_{s}$, for sector $s$ is then the \textbf{concatenation} of $o_{s,w}$ over all $w$ in all codebooks $c_s \in \cexpert$.

This observation can be enriched with additional features, such as beam alignment feedback, SSB-specific metrics, or other relevant KPIs, enabling the policy to leverage richer information and more accurately predict the codebook configuration that maximizes the chosen reward. Note that the size of the observations is fixed, as the codebooks have a fixed size $n$.
\item \textbf{Rewards:} Achieved coverage, defined as the number of  successfully served UEs. Equation~\eqref{eqn: separate objective} shows how the reward function can be separated across sectors.
\end{itemize}

This POMDP captures the sequential, combinatorial, and uncertain nature of codebook design. Our RL solution learns adaptive policies from partial observations, enabling scalable, dynamic SSB beam management.

\subsection{Neural Network Architecture}\label{ssec:NNarchitecture}

Building on the POMDP formulation, we implement a stochastic policy $p(a\mid o)$ that, given an observation $o_s$, tries to assign high probabilities $p$ to SSB codebooks with high user coverage. For a sector $s$, the full codebook selection is factorized sequentially via the chain rule:
\begin{equation}
p(a_s \mid o_s) = \prod_{i=1}^{n} p\big( a_{s,i} \mid a_{s,<i}, o_s \big),
\end{equation}
$a_{s,i}\in\mathcal B$ is the $i$-th selected beam for sector $s$. This allows the policy to condition each beam selection $a_{s,i}$ on both the observation $o_s$ and previously chosen beams $a_{s,<i}$, similarly to Pointer Networks \cite{vinyals2015pointer}.

Following \cite{bello2016neural}, we use an actor-critic architecture. Both actor and critic networks process the observation vector: the critic estimates the expected reward, and the actor samples $n$ SSBs without replacement to form the codebook.


The training objective of the actor is to maximize the expected coverage for a given observation $o_s$. We define a loss function $L$ as the negative of the coverage achieved by the codebook chosen:
\begin{equation}
L(a_s\mid o_s) = -\sum_{u: s^*_u=s}\mathbbm{1}\!\left\{ P_u^{(s^*_u,w^*_u)} \geq \tau\right\} = -\sum_{w\in a_s}o_{s,w}
\end{equation}
In this formulation, minimizing the expected loss is equivalent to maximizing the expected coverage.
\begin{equation}
J(\thetaa \mid o) = \mathbb{E}_{a \sim p_{\thetaa}(\cdot \mid o)} \big[ L(a \mid o) \big]
\end{equation}
 Here, $\thetaa$ denotes the actor's neural network parameters, and the expectation is taken over codebooks sampled from the policy distribution $p_{\thetaa}(a \mid o)$. Using this convention, standard gradient-based optimization methods can be applied, and the policy is updated using a mini-batch of $K$ samples via the REINFORCE algorithm as:

\begin{align}
\nabla_{\thetaa} J(\thetaa \mid o) 
&\approx \frac{1}{K} \sum_{k=1}^{K} 
\Big( L(a^{(k)} \mid o^{(k)}) - \beta_{\thetac}(o^{(k)}) \Big) \nonumber \\
&\quad \times \nabla_{\thetaa} \log p_{\thetaa}(a^{(k)} \mid o^{(k)}).
\end{align}
Here, $\beta_{\thetac}(o^{(k)})$ is the baseline predicted by the critic. This ensures that minimizing the loss directly leads to learning policies that select codebooks maximizing coverage.

The critic is updated over the same mini-batch by minimizing the mean squared error between its predicted baseline and the observed reward. This helps stabilize training and reduces the variance of the gradient.
\begin{equation}
\mathcal L(\theta_\text{critic}) = \frac{1}{K} \sum_{k=1}^{K} \big\| \beta_{\theta_\text{critic}}(o^{(k)}) - L(a^{(k)} \mid o^{(k)}) \big\|_2^2.
\end{equation}

Having described the model architecture and optimization objectives, we now outline the training procedure for the proposed solution and explain how the resulting policy is deployed in practice.

\subsection{Training and Deployment}
The proposed approach operates in two modes: learning and deployment. During learning, the policy and value networks are trained using the actor–critic formulation described above in a simulated multi-cell environment modeling user distributions, traffic, and channel conditions. Each episode spans $|\cexpert|+1$ consecutive CSs: in the first $|\cexpert|$ CSs, each cell deploys an expert codebook and collects feedback, assuming UEs are initially unassociated. At CS $|\cexpert|+1$, each agent selects a codebook ($n$ out of $m$ SSBs) and receives the achieved coverage as reward.

Training proceeds over multiple iterations with mini-batches of $K$ episodes, applying observation normalization and advantage clamping for stability. Training stops when coverage converges or after a fixed maximum number of iterations.

After training, each cell runs an agent to select its codebook. Expert codebooks are deployed for multiple CS periods to allow user reconnection and ensure reliable feedback. The agent then selects the optimal $n$ SSBs for the next deployment window, with sweeping and reconfiguration intervals set empirically.

\section{Performance Evaluation} \label{sec:perf}
\subsection{Experimental setup}
\begin{figure}[h!]
    \centering
    \includegraphics[width=\linewidth]{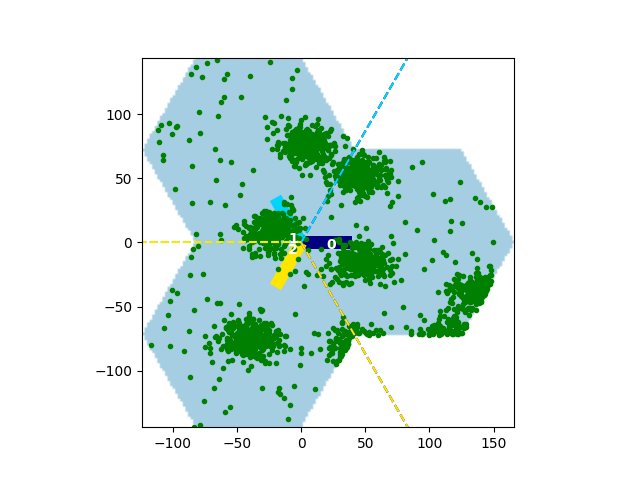}
        \caption{Example of generated environment with inhomogeneous UEs' deployment.}

    \label{fig:env}
\end{figure}

\textbf{Simulator}: Experiments use a proprietary mobile network simulator with one base station comprising three cells in the mmWave band. UEs are distributed inhomogeneously in each experiment instance, with Gaussian clusters modeling dense hotspots and up to $80\%$ of UEs are indoors (Fig.~\ref{fig:env}). The SSB pool $\mathcal B$ contains 144 expert-designed beams. To construct the observation $o_s$, each cell, $s$, initially performs the first two IA cycles using two expert codebooks of 24 beams each ($c_1$ and $c_2$, corresponding to the first 48 beams in $\mathcal B$) before designing the new codebook. The three sectors select the same expert codebook simultaneously, i.e., if one cell uses $c_1$, the others do as well. Codebook $c_1$ uses narrow beams for distant UEs, while $c_2$ uses broader beams for nearby UEs. Training uses randomly generated instances of the environment to improve generalization. The critic is a three-layer feedforward network estimating expected rewards, and the actor is a four-layer feedforward network with softmax output. As explained in Section \ref{ssec:NNarchitecture}, at each step the actor selects one beam according to this output probability distribution and then removes the beam from the possible actions for the remaining steps. The parameters of the actor and the critic are summarized in Table \ref{tab:actor_critic_params}.

\begin{table}[ht]
\centering
\caption{Actor and Critic Network Parameters}
\label{tab:actor_critic_params}
\begin{tabular}{ l c c }
\hline
\textbf{Parameter} & \textbf{Actor} & \textbf{Critic} \\
\hline\hline
Layer 1 Output Size & 512 & 256 \\ \hline
Layer 2 Output Size & 512 & 128 \\ \hline
Layer 3 Output Size & 256 & 1   \\ \hline
Layer 4 Output Size & 144 & --  \\ \hline
\end{tabular}
\end{table}

\begin{table}[h!]
\centering
\caption{Simulation Environment Parameters}
\label{tab:sim_param}
\begin{tabular}{lc}
\hline
\textbf{Parameter} & \textbf{Value} \\ \hline
Number of UEs & 2000 \\
Traffic volume bits/sec& $3\times10^{8}$ \\
Learning rate  & $10^{-3}$ \\
Frequency & 28 GHz \\
Number of iterations for training & $15\times10^{3}$ \\
Number of expert codebooks  & $2$ \\
$m/n$ & $144/24$ \\
Mini-batch size $K$& $36$ \\
\hline
\end{tabular}
\label{tab:results_summary}
\end{table}

\textbf{Baselines}: After training, we compare the performance of the RL-based codebook, referred to as the \emph{Neural Codebook}, against four baselines:
\begin{itemize}
\item \textbf{Expert codebooks $c_i$:} All three sectors use the same expert-designed codebook, $c_1$ or $c_2$, during CS. 
\item \textbf{Max of Experts:} In each environment, every cell sequentially sweeps $c_1$ and $c_2$ and selects the codebook that maximizes UE coverage.
\item \textbf{Greedy codebook:} Each cell performs CS with $c_1$ and $c_2$, ranks the 48 SSBs by UE coverage, and selects the top 24 beams to form a new codebook for CS.
\item \textbf{Random codebook:} Each cell randomly selects 24 SSBs from the 144 beams in $\mathcal B$ to construct a codebook for CS.
\end{itemize}
To evaluate performance, we generate $200$ independent environment instances for an inter-site distance (ISD) of $200$ m and another $200$ instances for an ISD of $400$ m, assessing the \emph{Neural Codebook} against all baselines in each case. Varying the ISD affects the traffic and UE distribution, as well as the maximum distance between UEs and the BS.

\subsection{Results \& Discussion}
We report in Table~\ref{tab:results_summary} the percentage of network deployment instances where the Neural Codebook and the baselines outperform in terms of coverage, successfully associating the highest number of UEs during CS. The results clearly highlight the superiority of the Neural Codebook, which achieves the best performance in the vast majority of scenarios: $82.9\%$ of instances for ISD $= 200$m and $90.45\%$ for ISD $= 400$m. In contrast, the expert-designed codebooks rarely outperform the Neural Codebook, with $c_1$ dominating only 6.3\% of instances at ISD $= 200$m and just $1.01\%$ at ISD $= 400$m, while $c_2$ achieves slightly better performance at larger ISD ($4.52\%$) compared to smaller ISD ($2.7\%$). This suggests that broader beams in $c_2$ provide some advantage when UEs are located farther from the BS. The greedy codebook performs best in only a small fraction of cases ($1.8\%$ at ISD $= 200$m and $2.01\%$ at ISD = $400$m). Its weakness stems from the fact that several SSBs across $c_1$ and $c_2$ may cover largely overlapping UE regions, leading to redundant selections and limiting diversity. As a result, simply ranking beams by coverage fails to construct a truly efficient codebook compared to the RL-driven adaptive design. Similarly, the random codebook occasionally outperforms structured approaches ($6.3\%$ at ISD $= 200$ m and $2.01\%$ at ISD $= 400 m$), but these cases are inconsistent and highlight the inefficiency of unstructured selection.

\begin{table}[h!]
\centering
\caption{Percentage of experiments in which each baseline achieves the best performance for two ISD values.}
\label{tab:results_summary}
\begin{tabular}{lcc}
\hline
\textbf{Baseline} & \textbf{ISD = 200 m (\%)} & \textbf{ISD = 400 m (\%)} \\ \hline
Neural Codebook   & 82.9  & 90.45 \\
$c_1$ & 6.3   & 1.01  \\
$c_2$ & 2.7   & 4.52  \\
Greedy Codebook   & 1.8   & 2.01  \\
Random Codebook   & 6.3   & 2.01  \\ \hline
\end{tabular}
\end{table}


Table~\ref{tab:ratio_connected_UEs} presents the fraction of connected UEs (with respect to the total UEs in the system) achieved by each baseline under the two ISD scenarios, averaged over the deployment instances used for evaluation. The second column shows the fraction of connected UEs at ISD = 200 m, while the third column reports the fractions of connected UEs at ISD = 400 m.
    The results clearly show that the Neural Codebook consistently outperforms all other baselines. At ISD = 200 m, it achieves a score of 0.454, exceeding the max-of-experts and greedy baselines by $2.5\%$. 
It also outperforms the random baseline by $4.6\%$. At ISD $= 400$ m, the Neural Codebook reaches $0.624$, exceeding the best-performing expert codebook by $2.0\%$, $c_1$ by $2.5\%$, $c_2$ by $6.4\%$, greedy by $3.7\%$, and random by $3.2\%$.
Increasing the ISD generally improves coverage ratios for all baselines due to more distributed UE locations. Despite this, the Neural Codebook consistently maintains a performance margin, highlighting its robustness and adaptability to varying traffic and channel conditions.

Overall, the Neural Codebook improves coverage by $2–6.4\%$ over all baselines, demonstrating a clear and consistent advantage across different network deployments.

\begin{table}[h!]
\centering
\caption{Mean fraction of connected UEs for every baseline for ISD = 200 m and ISD = 400 m.}
\label{tab:ratio_connected_UEs}
\begin{tabular}{lcc}
\hline
\textbf{Baseline} & \textbf{ISD = 200} & \textbf{ISD = 400} \\ \hline
Neural Codebook   & 0.454 & 0.624  \\
$c_1$ &  0.414 &  0.599\\
$c_2$ &  0.423 & 0.56\\
Max of Experts    & 0.429 & 0.604 \\
Greedy Codebook   & 0.429 & 0.587 \\
Random Codebook   & 0.408 & 0.592 \\
\hline
\end{tabular}
\end{table}

\begin{figure}[!t]
    \centering
    \begin{subfigure}[b]{0.48\linewidth}
        \centering
        \includegraphics[width=\linewidth]{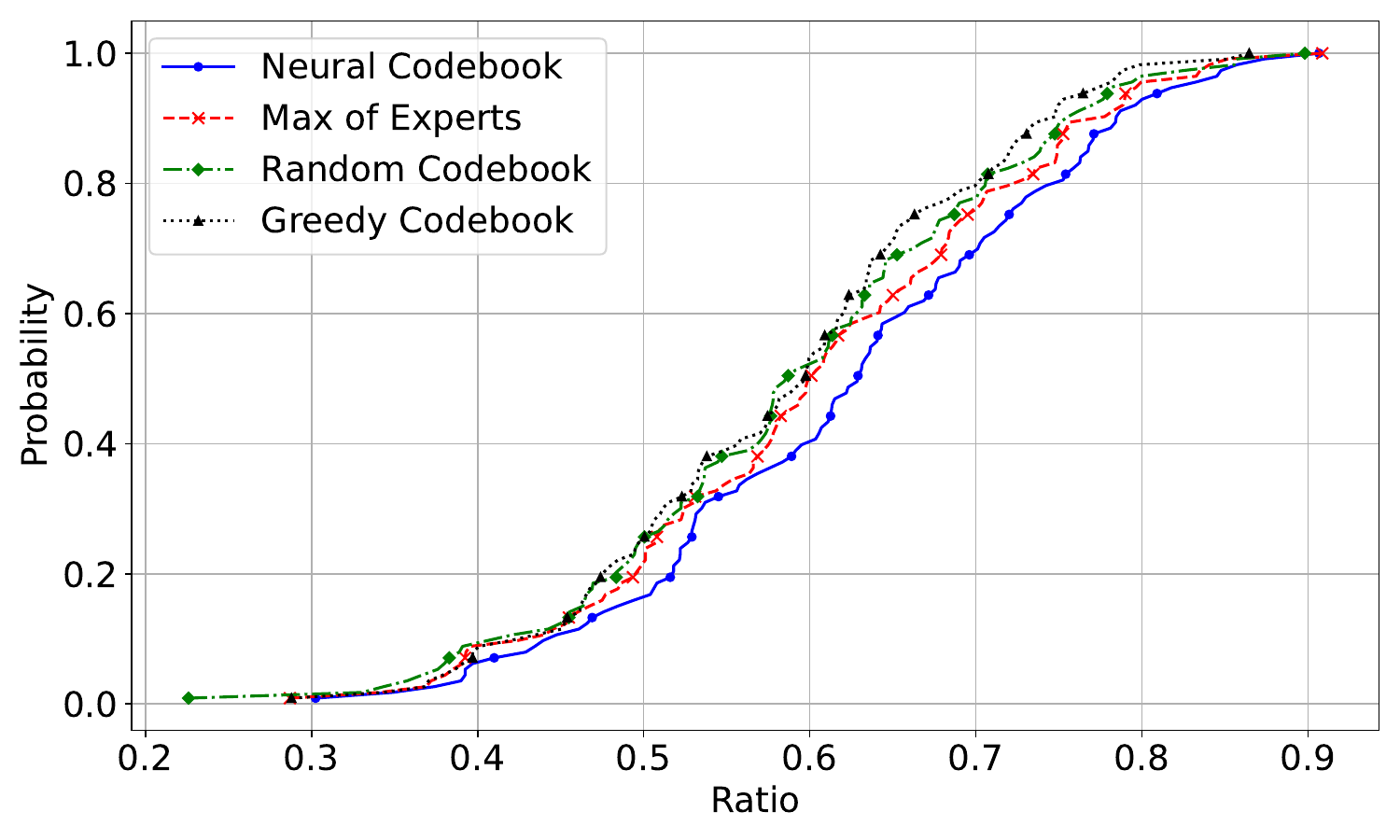}
        \caption{ISD $=200$m}
        \label{fig:CDF_ISD200}
    \end{subfigure}
    \hfill
    \begin{subfigure}[b]{0.48\linewidth}
        \centering
        \includegraphics[width=\linewidth]{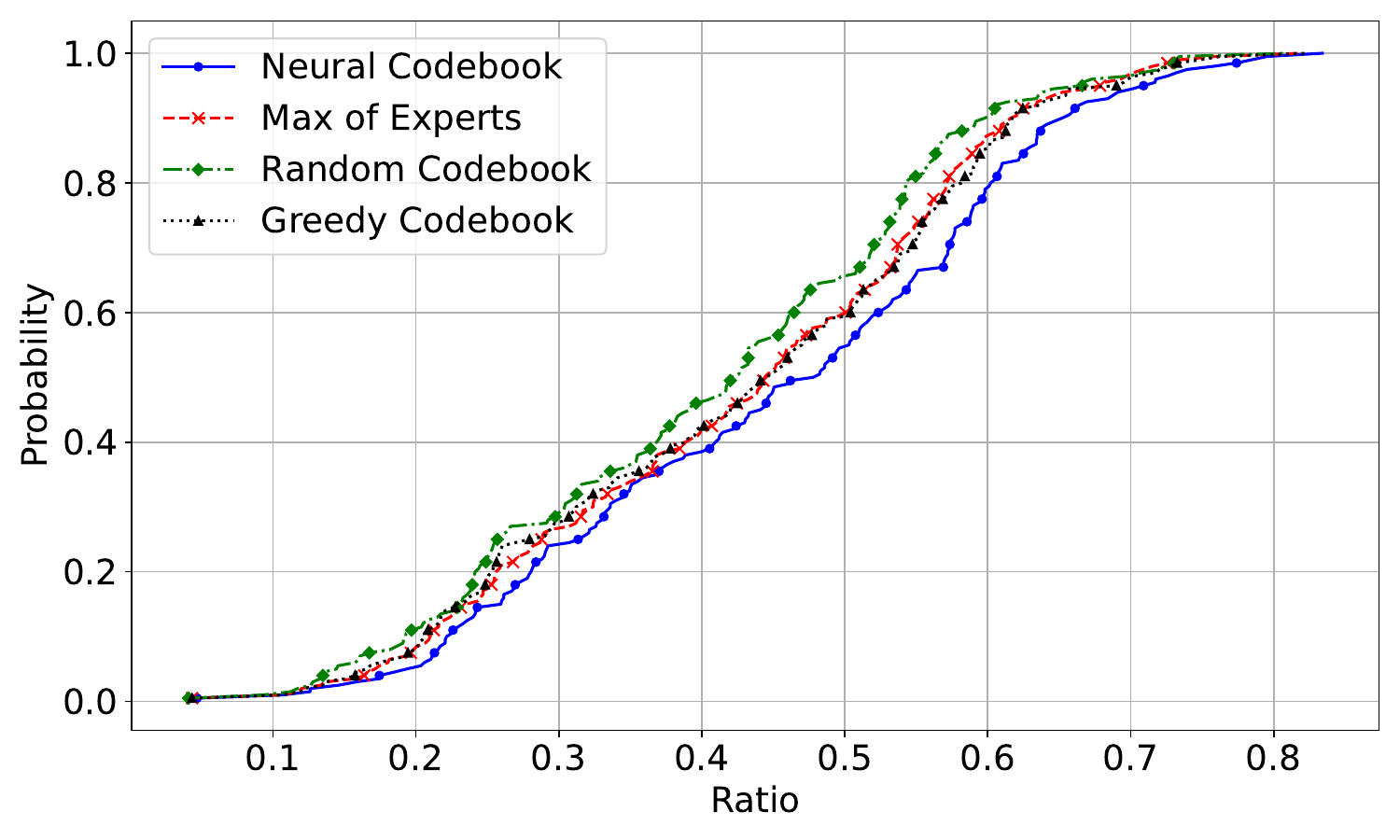}
        \caption{ISD $=400$m}
        \label{fig:CDF_ISD400}
    \end{subfigure}
    \caption{CDF of the fraction of connected/covered UEs.}
    \label{fig:CDF_all}
\end{figure}

 We complement the result in Table \ref{tab:ratio_connected_UEs}, by Figure \ref{fig:CDF_all} that presents The CDF of connected UEs for ISD = $200$m and ISD = $400$m. The two Figures confirm the trends in Table~\ref{tab:results_summary}. The Neural Codebook consistently achieves higher coverage across network instances, with its curve shifted to the right compared to all baselines. This indicates both higher average performance and greater reliability. In contrast, expert, greedy, and random codebooks have lower coverage, with CDFs concentrated toward smaller connected UE ratios.
 
\begin{figure}[h!]
    \centering
    \includegraphics[width=\linewidth]{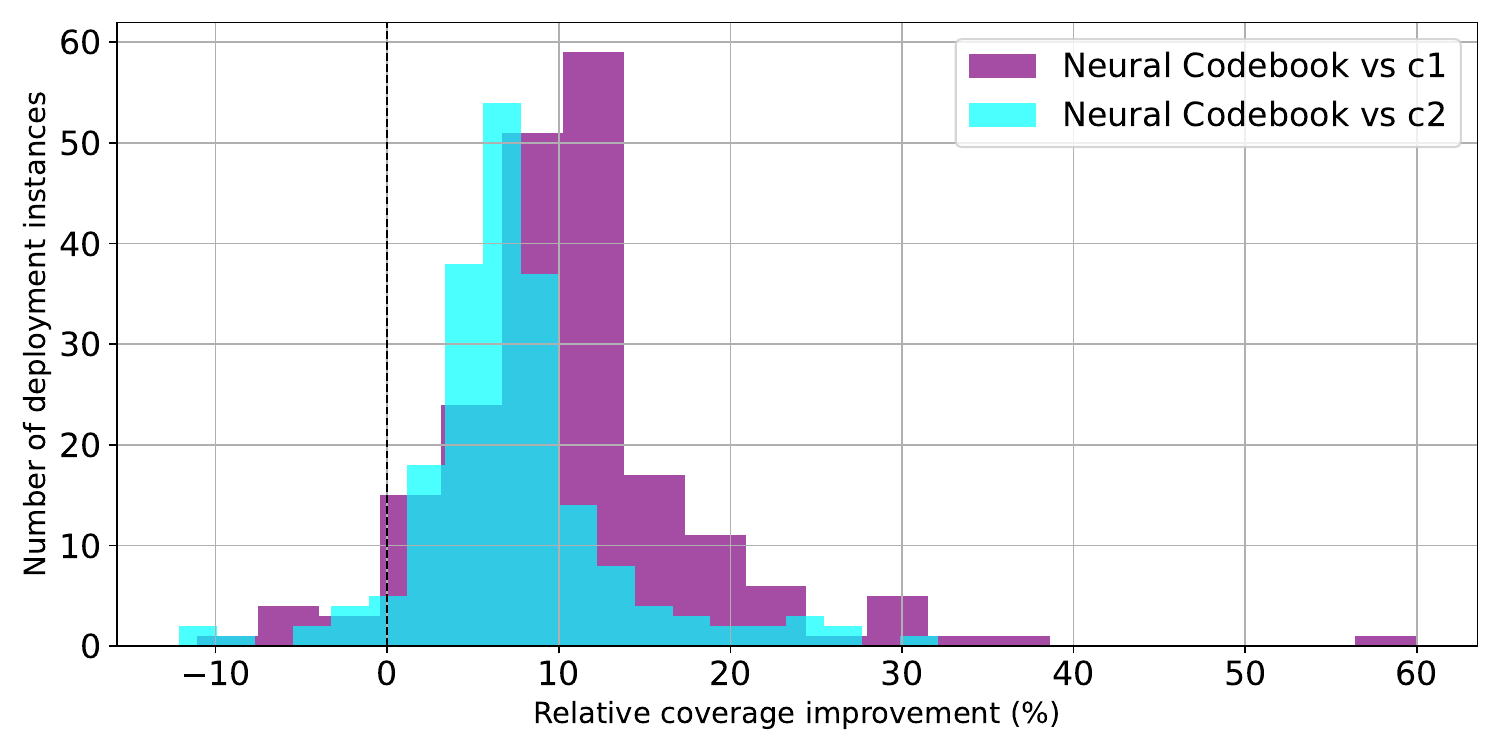}
        \caption{Relative improvement (\%) in the number of connected devices with the Neural Codebook over expert codebooks $c_1$ (purple) and $c_2$ (cyan).}

    \label{fig:histogram}
\end{figure}

The histogram in Fig.~\ref{fig:histogram} shows the distribution of relative coverage improvements achieved by the Neural Codebook compared to the two baselines, $c_1$ and $c_2$. In most deployment instances, the Neural Codebook provides positive gains, clustering around a $10\%$ improvement over $c_1$ and $7\%$ over $c_2$, with only a few rare degradations. Quantitatively, it outperforms $c_1$ in $192$ out of $200$ scenarios ($96\%$) and $c_2$ in $188$ out of $200$ scenarios ($94\%$). In the best cases, the improvement reaches $60\%$ over $c_1$ and $32.1\%$ over $c_2$. On average, the relative improvement is $10.8\%$ versus $c_1$ and $7.4\%$ versus $c_2$, demonstrating the robustness and effectiveness of the Neural Codebook across diverse deployment conditions.

\begin{figure}[!t]
    \centering
    \begin{subfigure}[b]{0.48\linewidth}
        \centering
        \includegraphics[width=\linewidth]{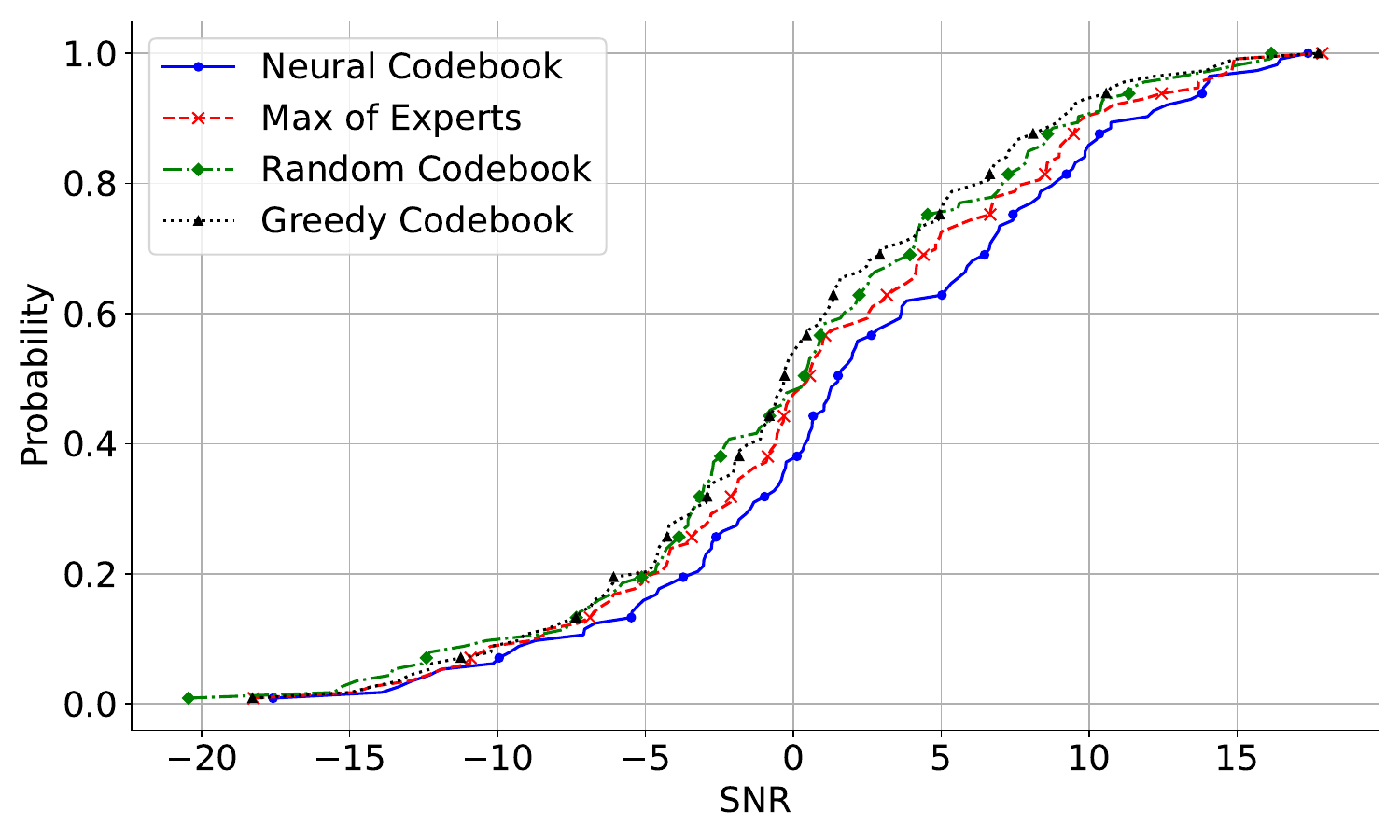}
        \caption{ISD $=200$m}
        \label{fig:TOPSNR_ISD200}
    \end{subfigure}
    \hfill
    \begin{subfigure}[b]{0.48\linewidth}
        \centering
        \includegraphics[width=\linewidth]
        {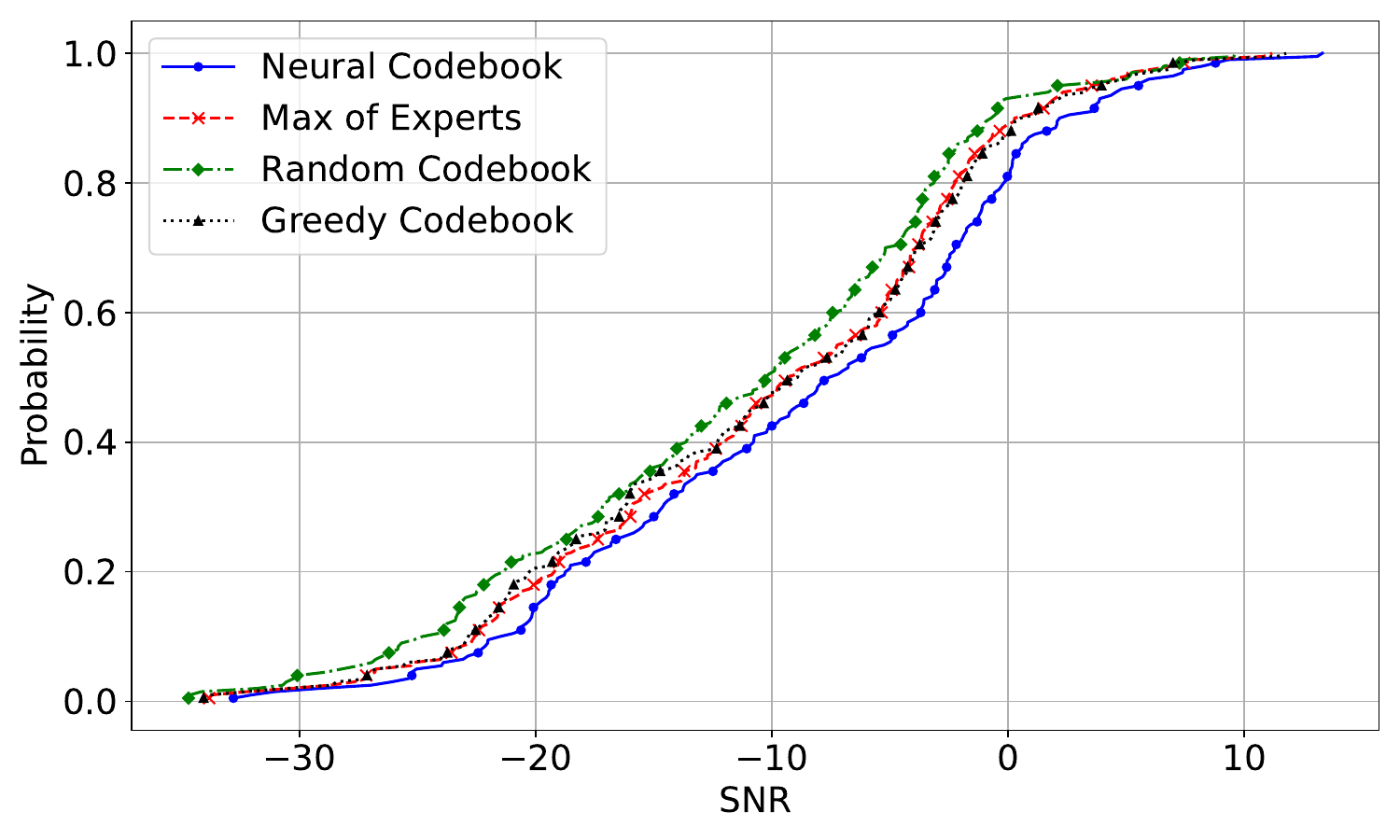}
        \caption{ISD $=400$m}
        \label{fig:TOPSNR_ISD400}
    \end{subfigure}
    \caption{Top $10\%$ rediscovered UEs' SSB SNR.}
    \label{fig:TOPSNR}
\end{figure}

In Figure~\ref{fig:TOPSNR}, we report the CDF of the average SSB's SNR for the top $10\%$ of deployed UEs. Although the RL agent was trained with a coverage-oriented reward (not explicitly optimizing SNR), the Neural Codebook still achieves slightly higher SNR values compared to all baselines. Its CDF is consistently shifted to the right, indicating that a larger fraction of deployments reach higher SNR levels. This suggests that the learned policy tends to select SSBs with higher gains as a byproduct of coverage maximization. The observed improvement, while modest, highlights the flexibility of our approach: by changing the reward to directly target SNR (or other KPIs), the RL framework could adapt its behavior to optimize for high-SNR UEs more aggressively. The Max of Experts and Greedy codebooks perform closely, whereas the Random codebook exhibits the largest spread, indicating less predictable performance.

Finally, Table \ref{tab:rediscovery} presents the Neural Codebook’s performance in terms of rediscovering UEs already served by at least one expert codebook (Union($c_1, c_2$)) and in identifying new UEs. The network successfully rediscovered $97.5\%$ of existing UEs for ISD = $200$m and $98.6\%$ for ISD = $400$m. Additionally, it discovered $9\%$ of new UEs at ISD = $200$m and $4\%$ at ISD = $400$m. These results demonstrate that the neural network effectively captures the coverage of the expert codebooks while providing a modest increase in overall UE connectivity.

\begin{table}[h!]
\centering
\caption{Rediscovery and newly discovered UEs ($\%$).}
\label{tab:rediscovery}
\begin{tabular}{lcc}
\hline
\textbf{Metric} & \textbf{ISD $= 200$ m} & \textbf{ISD = $400$ m} \\ \hline
Rediscovered UEs & $97.53\%$ & $98.6\%$ \\
Newly discovered UEs & $9\%$ & $4\%$ \\
\hline
\end{tabular}
\end{table}

Across diverse deployments, the Neural Codebook outperforms expert, greedy, and random baselines, improving coverage by \(2\!-\!6.4\%\) and reducing poor-performance cases. These results underscore the limits of heuristics and the value of data-driven, adaptive optimization.

\section{Conclusion} \label{sec:conc}
In this paper, we addressed the design of static, data-driven Synchronization Signal Block codebooks for initial access in 5G mmWave networks. We proposed a Reinforcement Learning-based framework that learns to construct an SSB codebook using feedback from expert-designed codebooks. The RL agent selects the most appropriate group of SSBs from a larger pool provided by radio experts, aiming to maximize initial access coverage. Unlike existing approaches, our method can adaptively redesign the codebook without retraining. Different from existing studies, we benchmarked the proposed solution against real expert codebooks as a strong baseline, and performance evaluations demonstrate that the RL-designed codebook consistently achieves higher UE coverage 
and improved overall network performance. These results highlight the potential of reinforcement learning for intelligent, adaptive beam management in mmWave networks, offering a practical and effective alternative to traditional codebook design strategies.

\bibliographystyle{IEEEtran}

\bibliography{IEEEabrv, SAP_bib}

\end{document}